\documentclass[11pt,a4paper]{article}

\usepackage{amsfonts}
\usepackage{color}
\usepackage{graphicx}

\title{\textbf{Non-topological kink scattering \\
		in a two-component scalar field theory model}}

\author{A. Alonso-Izquierdo$^{(a,b)}$
\\ $^{(a)}$ Departamento de Matematica Aplicada, Universidad de Salamanca, \\ Casas del Parque 2, 37008 - Salamanca, Spain \\
$^{(b)}$ IUFFyM, Universidad de Salamanca, \\ Plaza de la Merced 1, 37008 - Salamanca, Spain }

\date{}

\begin{document}

\maketitle

\begin{abstract}
In this paper the scattering between the non-topological kinks arising in a family of two-component scalar field theory models is analyzed. A winding charge is carried by these defects. As a consequence, two different classes of kink scattering processes emerge: (1) collisions between kinks that carry the same winding number and (2) scattering events between kinks with opposite winding number. The variety of scattering channels is very rich and it strongly depends on the collision velocity and the model parameter. For the first type of events, four distinct scattering channels are found: \textit{kink reflection} (kinks collide and bounce back), \textit{one-kink (partial) annihilation} (the two non-topological kinks collide causing the annihilation of one half of each kink and the subsequent recombination of the other two halves, giving rise to a new non-topological kink with the opposite winding charge), \textit{winding flip kink reflection} (kinks collide and emerge with the opposite winding charge) and \textit{total kink annihilation} (kinks collide and decay to the vacuum configuration). For the second type of events, the scattering channels comprise \textit{bion formation} (kink and antikink form a long-living bound state), \textit{kink-antikink passage} (kinks collide and pass each other) and \textit{kink-antikink annihilation} (kink and antikink collide and decay to the vacuum configuration). 
\end{abstract}

\section{Introduction}

Kinks are solutions in scalar field theory models whose energy density is localized and, thus, they can be interpreted as extended particles in some physical systems. The characteristics of these solutions have been extensively exploited to explain some non-linear phenomena arising in diverse branches of Physics. To mention just a few examples: signal transmission in optical fibers \cite{Mollenauer2006, Schneider2004, Agrawall1995}, the analysis of some features of DNA \cite{Yakushevich2004} and other molecular systems \cite{Davydov1985, Bazeia1999}, the description of magnetic flux quanta (fluxons) in Josephson junctions \cite{Ustinov1998, Kivshar1989}, the properties of some materials in Condensed Matter \cite{Bishop1980}, the behavior of the Early Universe \cite{Vilenkin1994, Vachaspati2006,Gani2018b}, etc. Obviously, kinks can move in the physical substrate involved in the previously mentioned applications, so the understanding of the collision processes between these traveling kinks is an essential issue. Indeed, kink scattering has been profusely studied both in Physics and Mathematics. In the case of scalar field theory models, such as those considered in this paper, kinks must comply with non-linear Klein-Gordon partial differential equations. In general, these equations lead to non-integrable systems. The sine-Gordon model constitutes one of the well-known exceptions. Curiously, kink scattering in these non-integrable systems exhibits a much richer behavior than that found for the integrable systems. This fact becomes apparent for the other paradigmatic member of the one-component scalar field theories, the $\phi^4$ model. The collision between kinks and antikinks in the $\phi^4$-model has been thoroughly studied in the seminal papers \cite{Sugiyama1979, Campbell1983, Anninos1991, Takyi2016}. In this case two possible scattering channels emerge: (a) \textit{bion formation} (kink and antikink collide and bounce back over and over emitting radiation in every impact) and (b) \textit{kink reflection} (kink and antikink collide and bounce back one or more times and finally escape with a certain separation velocity $v_f$). The appearance of these possible final scenarios critically depends on the collision velocity. One of the most remarkable aspects in this model is the presence of the \textit{resonant energy transfer mechanism}, which allows an energy exchange between the zero and vibrational kink modes, see \cite{Campbell1983}. As a consequence, the transition between the two previously mentioned regimes is characterized by a sequence of velocity windows with a fractal structure, where the previous regimes are interlaced and the kinks must collide a finite number of times before definitely escaping. An analytical explanation of this phenomenon using collective coordinates is given in References \cite{Goodman2005, Goodman2008, Goodman2007}. The existence of resonant energy transfer mechanisms has also been investigated in other one-component scalar field theory models, such as in the double sine-Gordon model \cite{Shiefman1979, Peyrard1983, Campbell1986, Gani1999, Malomed1989, Gani2018}, in deformed $\phi^4$ models \cite{Simas2016,Gomes2018,Bazeia2017b,Bazeia2017a, Bazeia2019}, in $\phi^6$-models \cite{Dorey2011, Romanczukiewicz2017, Weigel2014, Gani2014, Bazeia2018b}, etc. Recently, it has been proved that this mechanism can be activated by quasi-normal modes \cite{Dorey2018, Romanczukiewicz2018}. Kink scattering has also been studied for models with polynomial potentials of higher degrees. Here, if the potentials have global minima of quartic or higher degree, kinks with power-law asymptotics arise in the model, which leads to long-range interactions between the kinks. The exploration of this type of interactions must be carefully addressed and, nowadays constitutes a topic of significant interest \cite{Belendryasova2017, Christov2018, Gani2015, Christov2018, Gomes2012, Radomskiy2017, Bazeia2018c, Bazeia2018d, Christov2018b, Manton2019}. The effect of defects, impurities or inhomogeneities on kink dynamics has been discussed in several models, see references \cite{Fei1992,Fei1992b, Goodman2002, Goodman2004, Malomed1985, Malomed1992, Javidan2006, Saadatmand2012,Saadatmand2013, Saadatmand2015,Saadatmand2018, Adam2018}. The simultaneous collision of $N$ kinks has also been recently studied for some of the previously mentioned systems, see \cite{Marjaneh2017, Marjaneh2017b, Marjaneh2017c}. Furthermore, the dynamics of kinks with excited internal vibrational eigenmodes is also a remarkable issue. The evolution of the wobbling kink in the $\phi^4$-model was initially studied by Getmanov \cite{Getmanov1976}. An analytical expression for the decay of the amplitude of the wobbling mode has been deduced by Barashenkov and Oxtoby using singular perturbation expansions, see \cite{Barashenkov2009,Barashenkov2018}. It is also noteworthy the appearance of spectral walls in models for which the bound modes disappear into the continuous spectrum when the kinks approach each other. This phenomenon has been studied in the $\phi^4$ model with a BPS preserving impurity by Adam, Oles, Romanczukiewicz and Wereszczynski in \cite{Adam2019}.

All the previously cited works concern with scalar field theories involving only one field. In this context, the kink orbit consists of a line segment embedded in the one-dimensional internal space $\mathbb{R}$. This fact strongly constrains the possible final scenarios in kink scattering processes. The collective coordinate method takes advantage of this behavior and a configuration with a finite number of freedom degrees can usually be employed to approximate the evolution of the kinks in this type of theories. The situation is drastically changed when models with two or more scalar fields are considered. To begin with, the identification of exact static kink solutions in this context is much more difficult than for one-component scalar field theories. 
Indeed, this topic has led to an active research area during the last decades since Rajaraman et al. drew attention to this class of problems \cite{Rajaraman1982}. Some models in this context have received great interest in the physical and mathematical literature due to the properties of its kink varieties. For instance, the MSTB model (named after Montonen, Sarker, Trullinger and Bishop) is the simplest generalization of the $\phi^4$ model to two-dimensional internal space preserving the existence of two vacua \cite{Montonen1976, Sarker1976}. The kink variety for this system can be analytically identified and, indeed, exhibits a very rich structure. It consists of a pair of stable two-component kinks, an unstable one-component kink and two one-parametric families of unstable non-topological kinks, all of them complying with an intriguing energy sum rule, see references \cite{Rajaraman1975, Currie1979, Rajaraman1979, Subbaswamy1980, Subbaswamy1981, Magyari1984, Ito1985, Ito1985b, Guilarte1987, Guilarte1988, Alonso1998}. The MSTB model has been generalized for internal spaces with three fields \cite{Alonso2000, Alonso2002c, Alonso2004} and for nonlinear massive Sigma models where the internal space is constrained to a sphere \cite{Alonso2008,Alonso2009,Alonso2010,Alonso2018c}. The cornerstone, which underlies the properties of the kink varieties in these models, is that the static field equations are Hamilton-Jacobi separable \cite{Alonso2008b}. This fact allows the identification of all the static solitary wave solutions. The kink variety has also been analytically identified for the so called BNRT model, which was initially introduced by Bazeia, Nascimento, Ribeiro and Toledo  \cite{Bazeia1997,Bazeia1995} and studied later by other authors \cite{Shifman1998, Alonso2002, Alonso2002d, Alonso2014} in different contexts. In this case, the potential can be written as half the square of the gradient of a superpotential, which leads to first order equations using a Bogomolny arrangement of the energy functional. The resolution of these equations shows that there exist one-parameter kink families for different topological sectors. Kink solutions have also been calculated for models coupling the $\phi^4$ and/or sine-Gordon models \cite{Katsura2014, Dias2007, Avelino2009}, models with a real scalar Higgs field and a scalar triplet field \cite{Gani2016}, models coupled to gravity in warped spacetimes \cite{Dewolfe2000, Campos2002, Bazeia2004, Bazeia2015}, scalar field theories possessing self-dual sectors \cite{Adam2013, Adam2016, Ferreira2019}, etc. In addition, some deformation procedures have been developed, which allows to obtain exact solutions of two-field models from one-field models, see \cite{Bazeia2013}. 

While kink dynamics in models with one field has been extensively studied for some decades, the kink dynamics in two-component scalar field theory models has begun to be studied recently. This background has been employed in some physical applications, such as the analysis of the dynamics of magnetic flux solitons (fluxons) in systems of Josephson junctions \cite{Yukon2015, Abdumalikov2004,Goldobin2000}, which is modeled by a pair of coupled sine-Gordon equations for the phases of the junctions. A complex structure of true and false vacua arises in the two-dimensional internal space $(\phi_1,\phi_2)$ for this model, which induces the existence of different species of topological and non-topological solitons. Domain wall dynamics has been used as a candidate to explain the dark energy of the Universe without recourse to a non-vanishing cosmological constant \cite{Ashcroft2016}. Alternatively, kink dynamics in different contexts has been analyzed in coupled two-component $\phi^4$ models \cite{Halavanau2012, Romanczukiewicz2008}, in the MSTB model \cite{Alonso2018, Alonso2018b} and in the BNRT model \cite{Alonso2017}.

In this paper we are interested in studying the kink scattering for a family of two-component scalar field theory models in (1+1)-Minkowskian space-time with potential term defined by the polynomial of fourth degree in the fields $U(\phi_1,\phi_2)= 2 (\phi_1^2+ \phi_2^2-1)^2 +\frac{1}{2} \phi_2^2 +\sigma^2 (2\phi_1^2 + 2\phi_2^2-\phi_1-1)^2$. The novel characteristic of this model is that it involves only one vacuum, so kink solutions (if they exist) must necessarily be non-topological. The model can be interpreted as a deformation of the MSTB model, where the $\mathbb{Z}_2$ symmetry defined by the $\phi_1\rightarrow -\phi_1$ transformation is broken. Obviously, the integrability of the static MSTB field equations is also lost. As a consequence, only one type of non-topological kinks emerges as solutions in the new model. Despite the non-topological nature of these defects, they are stable with respect to small fluctuations if the model parameter $\sigma$ is large enough, specifically greater than approximately 0.71. The solutions asymptotically start and end at the only vacuum of the model traversing a loop orbit that surrounds a potential peak. This means that the extended particles described by these non-topological defects carry a winding charge determined by the orientation in which the closed kink trajectory is described. This fact implies that the two-body non-topological kink scattering processes must be classified into two categories: the first one comprises the collision events between extended particles that carry the same winding number, and the second one involves particles with opposite winding number. As we will see later, a very rich variety of scattering channels emerges, whose presence strongly depends on the collision velocity $v_0$ and the model parameter $\sigma$. Indeed, from our point of view some of the final scenarios become unexpected. 
For example, for the first type of events, the scattering channels are listed as follows: \textit{kink reflection} (kinks collide and bounce back), \textit{one-kink (partial) annihilation} (the two non-topological kinks collide causing the annihilation of one half of each kink and the subsequent recombination of the other two halves, giving rise to a new non-topological kink with the opposite winding charge), \textit{winding flip kink reflection} (kinks collide and emerge with the opposite winding charge) and \textit{total kink annihilation} (kinks collide and decay to the vacuum configuration). Furthermore, for the second class of events, the scattering channels comprise \textit{bion formation} (where kink and antikink form a long-living kink-antikink bound state), \textit{kink-antikink passage} (kinks collide and pass each other) and \textit{kink-antikink annihilation} (kink and antikink collide and decay to the vacuum configuration).

The organization of this paper is as follows: in Section 2 the model is introduced and the non-topological kinks describing the extended particles in this system are analytically identified. The linear stability study of these solutions is also addressed by proving that the non-topological kinks are stable for model parameters greater than 0.71 when small fluctuations are applied. Section 3 is devoted to the kink scattering analysis in this model. Firstly, a general description of the context is provided. The scattering events associated with the collision between two extended particles with the same winding charge are discussed in Section 3.1, whereas those with opposite winding charge are dealt with in Section 3.2. Finally, some conclusions are drawn in Section 4.

\section{The model}

\begin{figure}[h]
	\centerline{\includegraphics[height=3.5cm]{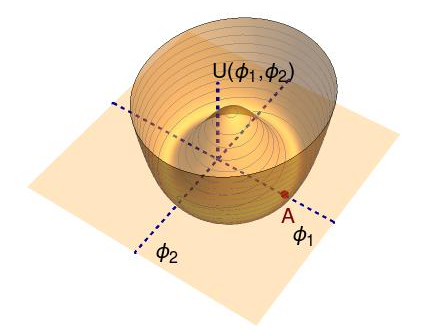} \hspace{1.5cm} \includegraphics[height=3.5cm]{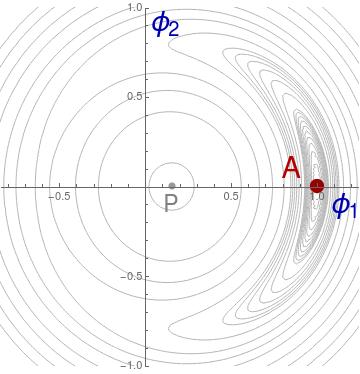}}
	\caption{\small Graphics of the potential term $U(\phi_1,\phi_2)$ represented by means of a 3D-plot (left) and a contour plot (right). The potential function $U(\phi_1,\phi_2)$ exhibits only one absolute minimum located at $A=(1,0)$.} \label{fig:potential}
\end{figure}

We shall deal with a (1+1)-dimensional two-coupled scalar field theory model whose dynamics is governed by the action
\begin{equation}
S=\int d^2 x \Big[ \frac{1}{2} \partial_\mu \phi_a \partial^\mu \phi_a - U(\phi_1,\phi_2) \Big] \hspace{0.4cm},
\label{action}
\end{equation}
where Einstein summation convention is assumed for $\mu=0,1$ and $a=1,2$. The Minkowski metric $g_{\mu \nu}$ has been chosen as $g_{00}=-g_{11}=1$ and $g_{12}=g_{21}=0$. The spacetime coordinates will be denoted as $x^0\equiv t$ and $x^1\equiv x$ from now on. The real scalar fields $\phi_a: \mathbb{R}^{1,1} \rightarrow \mathbb{R}$ with $a=1,2$ and the spacetime coordinates are assumed to be dimensionless in the expression (\ref{action}). In this paper kink dynamics ruled by the presence of the particular potential term
\begin{equation}
U(\phi_1,\phi_2)=2 (\phi_1^2+ \phi_2^2-1)^2 +\frac{1}{2} \phi_2^2 +\sigma^2 (2\phi_1^2 + 2\phi_2^2-\phi_1-1)^2  \label{potential}
\end{equation}
will be studied. This expression is a non-negative polynomial function of fourth degree in the field components $\phi_1$ and $\phi_2$, whose vacuum set ${\cal M}$ (formed by the absolute minima of the potential $U(\phi_1,\phi_2)$) consists of only one element
\begin{equation}
{\cal M} = \{ A =( 1,0) \} \hspace{0.5cm} . \label{vacua}
\end{equation}
The potential term (\ref{potential}) has been plotted in Figure \ref{fig:potential} by using a 3D-graphics (left) and a contour plot (right). It can be observed that $U(\phi_1,\phi_2)$ vanishes only at the minimum $A$ and that a local maximum is located at the point
\[
P\equiv \left(\frac{\sqrt{4+8\sigma^2+9\sigma^4} -(2+\sigma^2)}{4(1+2\sigma^2)},0\right) \hspace{0.5cm} .
\]
It is noteworthy that the action functional (\ref{action}) is invariant by the symmetry group $\mathbb{G}=\{e\}\times \mathbb{Z}_2$ generated by the transformation $\pi_2:(\phi_1,\phi_2)\mapsto (\phi_1,-\phi_2)$ but not by the transformation $\pi_1:(\phi_1,\phi_2)\mapsto (-\phi_1,\phi_2)$. This is due to the linear dependence on $\phi_1$ of the last summand in (\ref{potential}) .

\noindent The Euler-Lagrange equations derived from the functional (\ref{action}) lead to the coupled non-linear Klein-Gordon equations
\begin{eqnarray}
\frac{\partial^2 \phi_1}{\partial t^2} - \frac{\partial^2 \phi_1}{\partial x^2} &=& - 8\, \phi_1 (\phi_1^2 + \phi_2^2-1) - 2 \,\sigma^2 (4\phi_1-1) (2\phi_1^2 + 2\phi_2^2 -1-\phi_1) \hspace{0.5cm} , \label{kleinequations}\\
\frac{\partial^2 \phi_2}{\partial t^2} - \frac{\partial^2 \phi_2}{\partial x^2} &=&  -8\, \phi_2(\phi_1^2 + \phi_2^2-1) - \phi_2 -8\,\sigma^2\, \phi_2 (2\phi_1^2+ 2\phi_2^2-\phi_1-1) \hspace{0.7cm} . \nonumber
\end{eqnarray}
The zero energy static homogeneous solutions of (\ref{kleinequations}) correspond to the elements of ${\cal M}$. Therefore, $A\equiv (1,0)$ is the only vacuum of the system. The second order small fluctuation operator valued on the point $A$,
\[
{\cal H}[A_\pm]=\left( \begin{array}{cc} -\frac{d^2}{dx^2} + 2(8+9\sigma^2)   & 0 \\ 0 & -\frac{d^2}{dx^2} + 1 \end{array} \right)
\]
provides us with insight in the linear stability of the vacuum $A$. Obviously, the spectrum of this matrix operator is positive, which guarantees that the vacuum solution $A$ is stable with respect to the application of small fluctuations.

\begin{figure}[h]
	\centerline{\includegraphics[height=2.8cm]{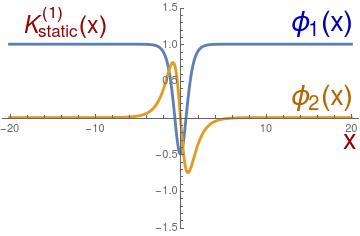} \hspace{0.8cm}  \includegraphics[height=2.8cm]{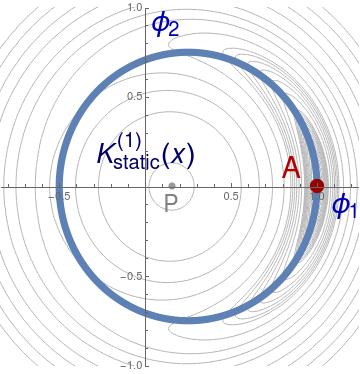} \hspace{0.8cm} \includegraphics[height=2.8cm]{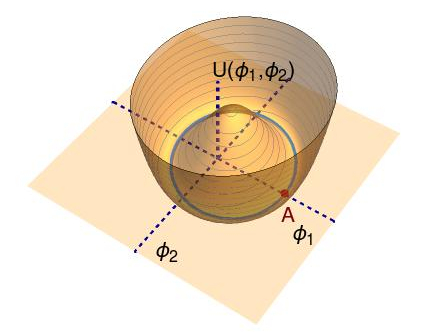} }
	\caption{\small Graphics of the profile (left) and orbit (middle and right) of the non-topological kinks $K^{(w)}_{\rm static}(x)$. A contour plot and a 3D graphics for the potential density $U(\phi_1,\phi_2)$ has been used in the second and third figures respectively.}  \label{fig:ntk}
\end{figure}

The spacetime translational symmetry of the action functional (\ref{action}) implies the conservation of the total energy
\begin{equation}
E[\Phi(x,t)]= \int_{-\infty}^\infty dx \,\, \varepsilon[\Phi(x,t)] \hspace{0.4cm}, \label{totalenergy}
\end{equation}
for any solution $\Phi(x,t)=(\phi_1(x,t),\phi_2(x,t))$ of the field equations (\ref{kleinequations}). The integrand $\varepsilon[\Phi(x,t)]$ of (\ref{totalenergy}) is the energy density
\begin{equation}
\varepsilon[\Phi(x,t)]= \frac{1}{2} \Big( \frac{\partial \phi_1}{\partial t} \Big)^2  + \frac{1}{2} \Big( \frac{\partial \phi_2}{\partial t} \Big)^2 +\frac{1}{2} \Big( \frac{\partial \phi_1}{\partial x} \Big)^2  +\frac{1}{2} \Big( \frac{\partial \phi_2}{\partial x} \Big)^2 + U(\phi_1(x,t),\phi_2(x,t)) \hspace{0.3cm}
\label{energydensity}
\end{equation}
of the solution $\Phi(x,t)$. The configuration space ${\cal C}$ in this context is defined as the set of maps $\Phi :\mathbb{R}^{1,1} \rightarrow \mathbb{R}\times \mathbb{R}$, whose total energy is finite, i.e., ${\cal C}=\{\Phi(x,t) \in \mathbb{R}\times \mathbb{R} : E[\Phi(x,t)]<+\infty \}$. From this definition, the elements of ${\cal C}$ must comply with the asymptotic conditions
\begin{equation}
\lim_{x\rightarrow \pm \infty} \frac{\partial \Phi(x,t)}{\partial t} = \lim_{x\rightarrow \pm \infty} \frac{\partial \Phi(x,t)}{\partial x} = 0 \hspace{0.5cm},\hspace{0.5cm} \lim_{x\rightarrow \pm \infty} \Phi(x,t) =  A \equiv (1,0) \hspace{0.3cm}. \label{asymptotic}
\end{equation}
As a consequence, solutions belonging to ${\cal C}$ must have a non-topological nature in our model.  They must asymptotically begin and end at the point $A$. However, this requirement does not exclude the existence of non-topological kinks. Indeed, a couple of static kinks (time-independent finite energy solutions of the field equations (\ref{kleinequations}) whose energy density (\ref{energydensity}) is localized) can be analytically identified. In particular, the expressions
\begin{equation}
K^{(w)}_{\rm static}(x)= \Big( 1-\frac{3}{2} \,{\rm sech}^2 \overline{x} \,\, , \,\, - (-1)^w \frac{3}{2} \, {\rm sech} \, \overline{x} \tanh \overline{x} \Big) \hspace{0.5cm} \mbox{with} \hspace{0.5cm} w=\pm 1 \hspace{0.5cm},
\label{ntk}
\end{equation}
describe non-topological kinks, which connect the vacuum $A$ with itself. Here, $\overline{x}=x-x_0$ with $x_0\in \mathbb{R}$ being the kink center. This solution traverses the circle centered at the point $(\frac{1}{4},0)$ with radius $\frac{3}{4}$ given by
\begin{equation}
\Big( \phi_1- \frac{1}{4} \Big)^2 + \phi_2^2 = \Big(\frac{3}{4} \Big)^2 \hspace{0.3cm} , \label{ntkorbit}
\end{equation}
surrounding the maximum potential peak placed at $P$, see Figure \ref{fig:ntk}. The parameter $w$ involved in (\ref{ntk}) is the winding number which sets the orientation in which the circle orbit (\ref{ntkorbit}) is traced. If the orbit (\ref{ntkorbit}) is traversed in a counter-clockwise direction the value of the winding number is positive, $w=1$, otherwise $w=-1$. Besides, the use of the mirror reflection in the space coordinate $\pi_x: x \mapsto -x$ allows us to write
\[
K^{(-1)}_{\rm static}(x) = \pi_x [K^{(1)}_{\rm static}(x)] = K^{(1)}_{\rm static}(-x) \hspace{0.5cm} .
\]
In this sense, $K^{(-1)}_{\rm static}(x)$ is the antikink of the $K^{(1)}_{\rm static}(x)$-kink. In particular, the term \textit{kinks} will be used in this paper to name solutions with positive winding number $w$, whereas those with $w=-1$ will be referred to as \textit{antikinks}.

\begin{figure}[h]
	\centerline{\includegraphics[height=3cm]{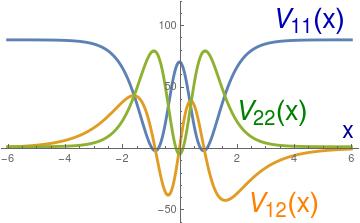} \hspace{1.5cm} \includegraphics[height=3cm]{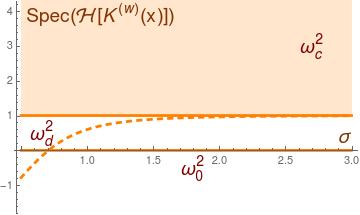} }
	\caption{\small Graphics of the potential well components of the small kink fluctuation operator (\ref{ntkhess}) (left) and dependence of its spectrum on the model parameter $\sigma$ (right).} \label{fig:ntkspectrum}
\end{figure}

\noindent The energy density of the solutions (\ref{ntk})
\begin{equation}
\varepsilon [K^{(w)}_{\rm static}(x)] = \frac{9}{4} \, {\rm sech}^2 \, \overline{x} \label{enerden}
\end{equation}
is localized around the kink center $x_0$ and it is independent of the winding number $w$. Therefore, these solutions can be interpreted as the basic extended particles of the system where the winding number $w=\pm 1$ plays the role of the particle charge. The integration of (\ref{enerden}) provides us with the total kink energy
\[
E[K^{(w)}_{\rm static}(x)]= \frac{9}{2}
\]
in dimensionless variables. The linear stability of the solutions (\ref{ntk}) can be studied by analyzing the spectral problem of the second order small kink fluctuation operator
\begin{equation}
{\cal H}[K^{(w)}_{\rm static}(x)] = \left( \begin{array}{cc} -\frac{d^2}{dx^2} + V_{11}(x) & V_{12}(x) \\ V_{12}(x) & -\frac{d^2}{dx^2} +V_{22}(x)  \end{array} \right) \label{ntkhess}
\end{equation}
where
\begin{eqnarray*}
V_{11}(x)&=& 2(8+9\sigma^2) - 18 (3+4 \sigma^2) \,{\rm sech}^2 \overline{x} + 36 (1+2 \sigma^2) \,{\rm sech}^4 \overline{x} \hspace{0.2cm} , \\
V_{12}(x)&=& -12 \, {\rm sech}\, \overline{x}  \tanh \overline{x}  \left( 2+3\sigma^2- 3(1+2\sigma^2) \, {\rm sech}^2 \overline{x}  \right)  \hspace{1.1cm} ,\\
V_{22}(x) &=& 1+ 6(5+12 \sigma^2) \, {\rm sech}^2 \overline{x}  - 36 (1+2\sigma^2)\, {\rm sech}^4  \overline{x}  \hspace{1.75cm} ,
\end{eqnarray*}
for our case. For the sake of illustration, the potential wells $V_{ij}(x)$ have been plotted in Figure \ref{fig:ntkspectrum} (left) for the particular coupling constant $\sigma=2.0$. The spectral problem associated with the matrix differential operator (\ref{ntkhess}) cannot be analytically solved so a numerical approach has been applied on this problem. In Figure \ref{fig:ntkspectrum} (right) the spectrum of this operator has been graphically represented for the model parameter $\sigma$ in the range $\sigma\in[0,3]$. It can be observed that a zero mode is always present, which can be easily identified as
\[
\psi_{\omega^2=0}(x) = \left(  \begin{array}{c} 3 \,{\rm sech}^2\,\overline{x} \tanh \overline{x} \\
\frac{3}{2} \, {\rm sech}\, \overline{x} \, (1-2\,{\rm sech}^2 \overline{x}) \end{array} \right) \hspace{0.4cm} .
\]
The spatial translational symmetry of equations (\ref{kleinequations}) underlies the existence of this eigenmode. Two continuous spectra emerge on the threshold values $1$ and $2(8+9\sigma^2)$, which correspond to the asymptotic values of the potential wells in the diagonal of (\ref{ntkhess}). Besides, the presence of a bound state has been numerically identified, whose eigenvalue is an increasing function of the model parameter $\sigma$. The function goes from the value $-1$ for $\sigma=0$ and enters into the continuous spectrum for large enough value of $\sigma$, see Figure \ref{fig:ntkspectrum} (left). This discrete eigenvalue is negative approximately for the parameter range $\sigma\in [0,0.71]$. In this regime, the non-topological kink (\ref{ntk}) is unstable and it will spontaneously decay to the vacuum configuration whenever a small fluctuation is applied. This process obviously involves the emission of a large amount of radiation. Recall that the non-topological kink circles the local maximum $P$ of the function $U(\phi_1,\phi_2)$. In this regime, the elastic forces associated with the quadratic terms involving the spatial derivatives in the functional (\ref{action}) overcome the potential barrier exerted by the substrate potential $U(\phi_1,\phi_2)$, see Figure \ref{fig:ntk} (right). Conversely, if the model parameter $\sigma$ is greater than $0.71$ the solution (\ref{ntk}) becomes linearly stable. Now, the effect of the potential barrier is stronger than the elastic forces, which avoids the kink decay.

In sum, the two-component scalar field theory model introduced in this Section involves the existence of one type of basic extended particles described by the non-topological kink solutions (\ref{ntk}). These particles have the same energy density distribution but they can carry different winding charges $w$, which take on the values $\pm 1$. These extended particles are stable when the coupling constant $\sigma$ is greater than $0.71$ with respect to the application of small linear fluctuations.

\section{Non-topological kink scattering}

The scattering processes between the extended particles described in the previous section are discussed now. The static solutions (\ref{ntk}) can be transformed into constant velocity traveling kinks by using a Lorentz boost, that is,
\[
K^{(w)}(x,t;v_0) = K_{\rm static}^{(w)} \Big( \frac{\overline{x}-v_0t}{\sqrt{1-v_0^2}} \Big)
\]
where the winding charge $w$ takes values on $\pm 1$. Taking into account the role of this particle charge, two different classes of two-body scattering events can be distinguished:

\begin{itemize}
	\item \textit{$K^{(w)}(x,t;v_0)-K^{(w)}(x,t;-v_0)$ scattering processes}. This class of  events consists on the collision between two non-topological kinks which carry the same winding charge $w$. In this case, the initial configuration for our scattering problem is given by the concatenation
	\begin{equation}
	K^{(w)}(x+x_0,t;v_0) \cup K^{(w)}(x-x_0,t;-v_0) \hspace{0.5cm} , \label{conca2}
	\end{equation}
	where $x_0$ is large enough to guarantee the smoothness of the configuration (\ref{conca2}). Therefore, (\ref{conca2}) describes two well-separated non-topological kinks (or antikinks), see Figure \ref{fig:concatenation} (left). The total winding number of (\ref{conca2}) is equal to 2 for a kink-kink configuration and $-2$ for an antikink-antikink arrangement. The profile of the first component in the concatenation (\ref{conca2}) is an even function with respect to the space variable $x$ whereas the second component is described by an odd function. The previously mentioned parity symmetries are preserved by the second order differential equations (\ref{kleinequations}), so the evolving field components must follow the same behavior. This fact restricts the number of possible final scenarios after the kink collision. The different events arising in this type of kink scattering processes will be described in Section 3.1.
	
	\begin{figure}[h]
		\centerline{\includegraphics[height=3cm]{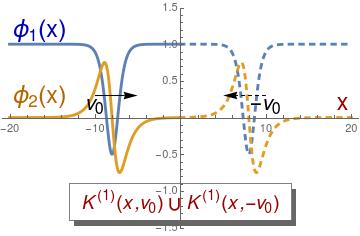} \hspace{1.5cm} \includegraphics[height=3cm]{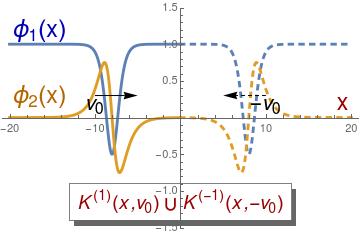} }
		\caption{\small Profile of the first and second component of the scalar field for the initial configuration (\ref{conca2}) associated to the $K^{(w)}(x,t;v_0)-K^{(w)}(x,t;-v_0)$ scattering processes (left) and for the arrangement  (\ref{conca3}) corresponding to the $K^{(w)}(x,t;v_0)-K^{(-w)}(x,t;-v_0)$ scattering events (right). } \label{fig:concatenation}
	\end{figure}

	\item \textit{$K^{(w)}(x,t;v_0)-K^{(-w)}(x,t;-v_0)$ scattering processes}. In this case, the scattering between two non-topological kinks with opposite winding charge is considered. Now the total winding number of the initial configuration vanishes. The initial configuration is characterized as
	\begin{equation}
	K^{(w)}(x+x_0,t;v_0) \cup K^{(-w)}(x-x_0,t;-v_0) \hspace{0.5cm} , \label{conca3}
	\end{equation}
	which represents a kink carrying winding number $w$ situated to the left of a kink with winding charge $-w$, see Figure \ref{fig:concatenation} (right). As usual, the parameter $x_0$ is chosen to be large enough so that the two single solutions in the kink-antikink (or antikink-kink) arrangement (\ref{conca3}) are well separated. Note that this type of events represents the collision between an extended particle and its own antiparticle. Here, the two field components involved in (\ref{conca3}) are even functions of the space variable $x$. This parity symmetry is preserved by the kink dynamics, which again restricts the number of possible scattering events. This context will be discussed in Section 3.2.
\end{itemize}

As we shall see later, the initial velocity $v_0$ plays an essential role in the kink scattering processes. Not only does this magnitude set the final separation velocity but also the nature of the resulting particles. In order to study the evolution of the initial configurations (\ref{conca2}) and (\ref{conca3}) dictated by the non-linear Klein-Gordon equations (\ref{kleinequations}), numerical analysis will be used. The numerical approach used in this paper follows the algorithm described in \cite{Kassam2005} by Kassam and Trefethen. This method is spectral in space and fourth order in time and was designed to solve the numerical instabilities of the exponential time-differencing Runge-Kutta method introduced in \cite{Cox2002}. As a complement to this numerical method, an energy conservative second-order finite difference algorithm \cite{Alonso2017}  implemented with Mur boundary conditions \cite{Mur1981}  has also been employed. The effect of radiation in the simulation is controlled by this algorithm because the linear plane waves are absorbed at the boundaries. The two previous numerical schemes provide similar results.

\subsection{$K^{(w)}(x)$-$K^{(w)}(x)$ kink scattering processes}

In this section the scattering between two extended particles which carry the same winding charge is discussed. With this purpose, the evolution of the initial configuration (\ref{conca2}), illustrated in Figure \ref{fig:concatenation} (left), has been numerically studied. Recall that parity symmetry of (\ref{conca2}) is preserved by the equations (\ref{kleinequations}), which implies that the first and the second scalar field components must be respectively described by even and odd functions for any time $t$. This fact restricts the possible final scenarios after the kink collision. In the following points the possible scattering processes are described and illustrated for several values of the model parameter $\sigma$ and the initial velocity $v_0$:

\begin{itemize}
	\item[\textbf{1.-}] \textit{Kink reflection:} In this type of processes the two kinks carrying winding charge $w$ approach each other, but when they are close enough they repel each other, bounce back and travel away, giving rise to an almost elastic scattering process. This class of events is symbolically represented as
	\[
	K^{(w)}(v_0) \cup K^{(w)}(-v_0) \rightarrow K^{(w)}(-v_0) \cup K^{(w)}(v_0) \hspace{0.5cm}.
	\]
	A particular kink collision belonging to this type of events has been illustrated in Figure \ref{fig:kinkkinkreflection} for coupling constant value $\sigma=2.0$ and initial velocity $v_0=0.4$.

	\begin{figure}[h]
		\centerline{\includegraphics[height=3cm]{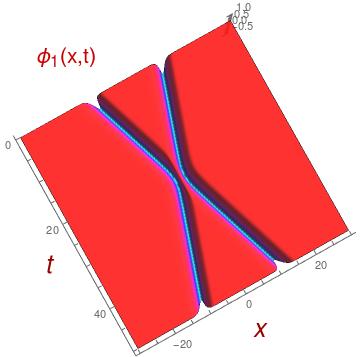} \hspace{1.5cm} \includegraphics[height=3cm]{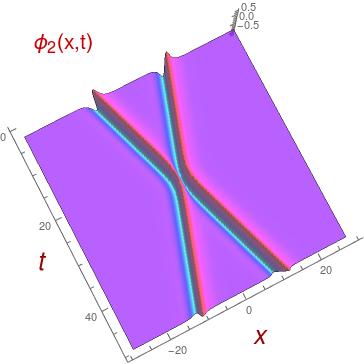}  \hspace{1.5cm} \includegraphics[height=3cm]{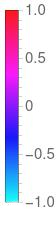}  }
		\caption{\small Evolution of the first and second scalar field components for a $K^{(w)}(x)$-$K^{(w)}(x)$ scattering process with impact velocity $v_0=0.4$ and model parameter $\sigma=2.0$: (\textit{Kink reflection}).} \label{fig:kinkkinkreflection}
	\end{figure}
	
	\item[\textbf{2.-}] \textit{One-kink annihilation:} This is the most curious scattering event arising in this model. In this case, the two kinks with winding charge $w$ approach each other, they collide and after the impact only one of the kinks survives, which remains pinned at the origin of the spatial axis. The process provokes a strong radiation emission coming from the one-kink annihilation. This type of events has been depicted in Figure \ref{fig:onekinkannihilation} for the case $\sigma=2.0$ with impact velocity $v_0=0.8$.
	
	\begin{figure}[h]
		\centerline{\includegraphics[height=3cm]{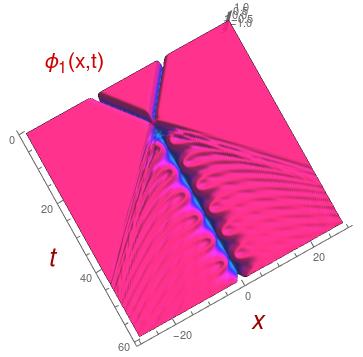} \hspace{1.5cm} \includegraphics[height=3cm]{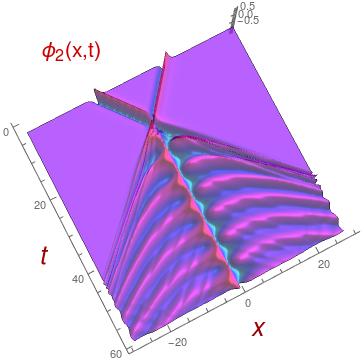} \hspace{1.5cm} \includegraphics[height=3cm]{colorbar}  }
		\caption{\small Evolution of the first and second scalar field components for a $K^{(w)}(x)$-$K^{(w)}(x)$ scattering process with impact velocity $v_0=0.8$ and model parameter $\sigma=2.0$.} \label{fig:onekinkannihilation}
	\end{figure}
	
	Taking into account the previously mentioned parity symmetry conservation, a thorough interpretation of these events is given as follows: the kinks approach each other, and when they are close enough the right-half of the right traveling kink interacts with the left-half of the left traveling kink, mutually annihilating and producing a strong radiation emission. The other two halves recombine each other giving rise to the surviving kink, which stays at the origin $x=0$. The most striking fact is that the winding number of the resulting extended particle is reversed. In other words, the initial configuration carrying a winding charge $2w$ evolves to a configuration with winding number $-w$. This behavior is outlined in the following scattering event
	\begin{equation}
	K^{(w)}(v_0) \cup K^{(w)}(-v_0) \rightarrow K^{(-w)*}(0) + \nu \label{partialannihilation} \hspace{0.5cm}.
	\end{equation}
	It has also been observed a strong excitation of the internal vibrational mode of the non-topological kink in this type of processes, which has been denoted by means of a asterisk superscript in (\ref{partialannihilation}). All the previous features of this partial annihilation can be better illustrated by showing the dynamics of the kink orbits for this event. In Figure \ref{fig:orbits11} the orbit of the evolving configuration starting from the initial arrangement (\ref{conca2}) has been depicted for several time instants. Note the complex behavior of the orbit evolution illustrated in this figure. Starting from a circle with winding number $2w$ (it is traversed twice), the loop formed by the left-half of the kink $K^{(w)}(v_0)$ and the right-half of the kink $K^{(w)}(-v_0)$ is deformed decreasing its size due to the orthogonal fluctuations provoked by the kink collision. This behavior can be observed for the times $t=12.0$ and $t=12.3$ in Figure \ref{fig:orbits11}. This finally leads to the annihilation of the previously mentioned kink halves. This can be visualized in the graphics for $t=12.8$ where the orbit already consists of a simple closed orbit. However, the induced fluctuations are so large that the other two kink halves also decay for a moment to the vacuum configuration (see orbits for $t=12.0$ and $t=13.4$) although later a new non-topological kink with the opposite winding number $-w$ emerges. Notice the large amount of radiation (plane waves) around the vacuum $A$ for $t=25.0$.

\begin{figure}[h]
	\centerline{\begin{tabular}{c}
			\includegraphics[height=2.cm]{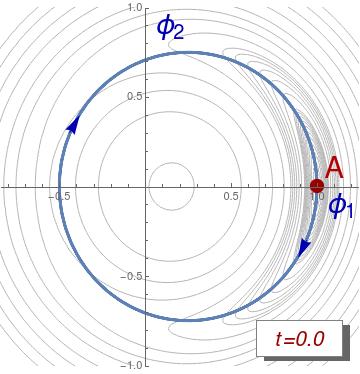}
			\includegraphics[height=2.cm]{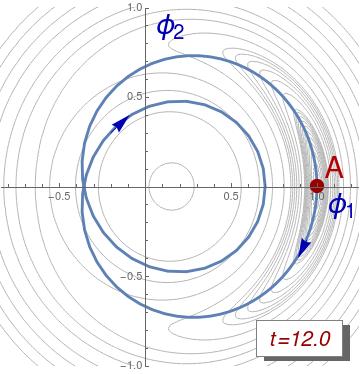}
			\includegraphics[height=2.cm]{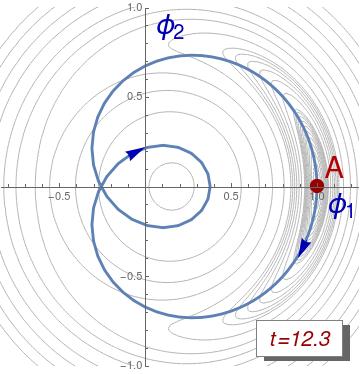}
			\includegraphics[height=2.cm]{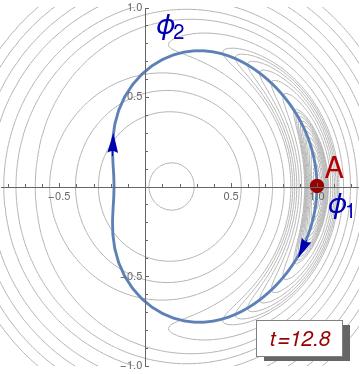}
			\includegraphics[height=2.cm]{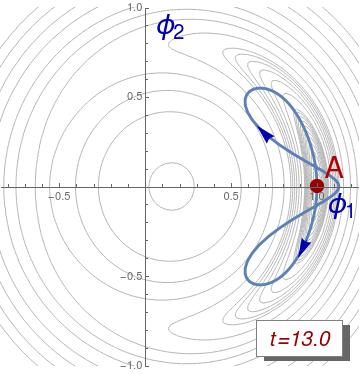}
			\includegraphics[height=2.cm]{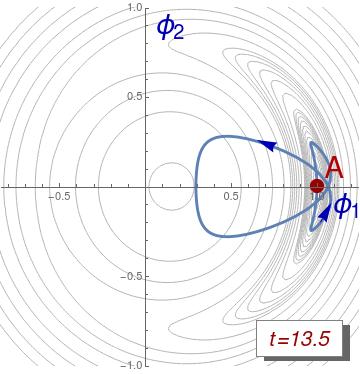}
			\includegraphics[height=2.cm]{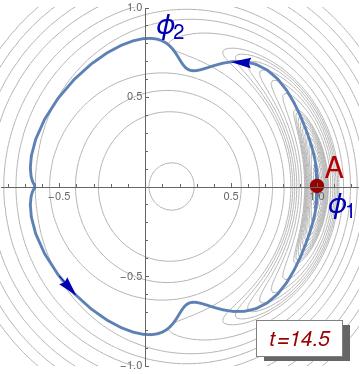}
			\includegraphics[height=2.cm]{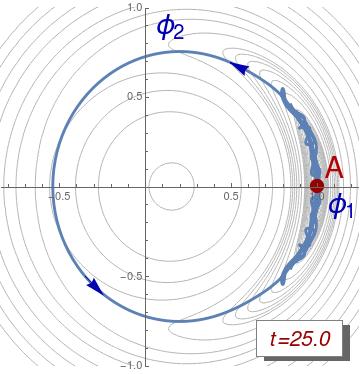} \end{tabular} }
	\caption{\small Evolution of the multi-kink orbit (\ref{conca2}) for the $K^{(w)}(x)-K^{(w)}(x)$-scattering process with impact velocity $v_0=0.8$ and model parameter $\sigma=2.0$ illustrated in Figure \ref{fig:onekinkannihilation}. Arrows on the curves indicate the direction in which the kink orbit is traversed.}
	\label{fig:orbits11}
\end{figure}

	\item[\textbf{3.-}] \textit{Winding flip kink reflection:} Other interesting event is given when the two kinks with the same winding charge $w$ approach each other, they collide, and after the impact they emerge with the winding charge opposite to which they initially carried. Thus, the initial configuration with a winding charge $2w$ evolves to a final configuration with winding number $-2w$. This pattern has been depicted in Figure \ref{fig:flipreflection} for the case $\sigma=0.95$ with impact velocity $v_0=0.68$. Note that continuous modes (radiation) of the multi-kink configuration are also excited in this process.

	\begin{figure}[h]
		\centerline{\includegraphics[height=3cm]{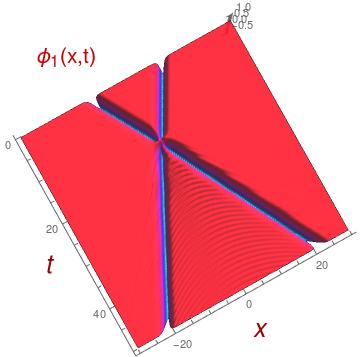} \hspace{1.5cm} \includegraphics[height=3cm]{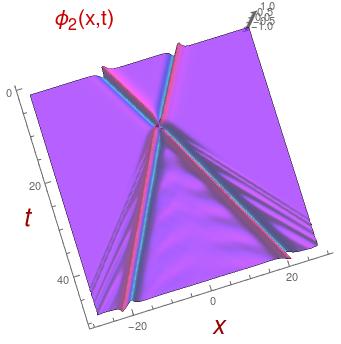} \hspace{1.5cm} \includegraphics[height=3cm]{colorbar}  }
		\caption{\small Evolution of the first and second scalar field components for a $K^{(w)}(x)$-$K^{(w)}(x)$ scattering process with impact velocity $v_0=0.68$ and model parameter $\sigma=0.95$.} \label{fig:flipreflection}
	\end{figure}

As in the previous case, the orbit dynamics in this case can help us to visualize this class of  scattering processes. In Figure \ref{fig:orbits22} the kink orbits for several relevant time instants have been plotted. Now, the orthogonal fluctuations induced by the kink impact deform the initial multikink configuration (\ref{conca2}), which is pushed towards the vacuum configuration (see the orbit sequence for $t=13.5$, $t=14.0$ and $t=14.8$) (this behavior suggests that the two original kinks are destroyed) but later, the accumulated kinetic energy causes the emergence of two new non-topological kinks although now with the opposite winding number $-w$ carried by the original kinks. This class of events can be represented as
\[
K^{(w)}(v_0) \cup K^{(w)}(-v_0) \rightarrow K^{(-w)}(-v_f) \cup K^{(-w)}(v_f)+ \nu \hspace{0.5cm} .
\]

\begin{figure}[h]
	\centerline{\begin{tabular}{c}
			\includegraphics[height=2.cm]{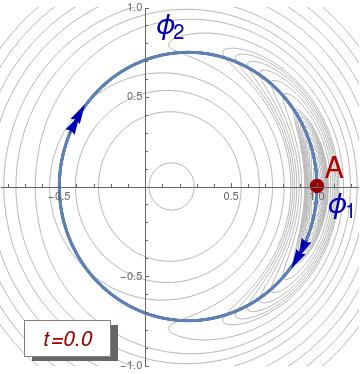}
			\includegraphics[height=2.cm]{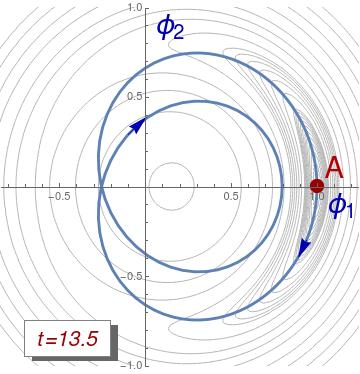}
			\includegraphics[height=2.cm]{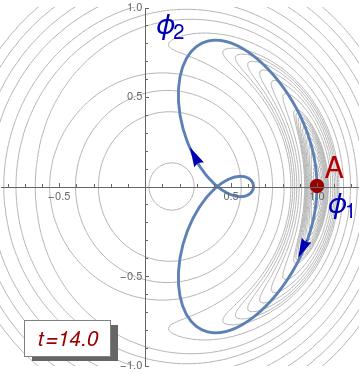}
			\includegraphics[height=2.cm]{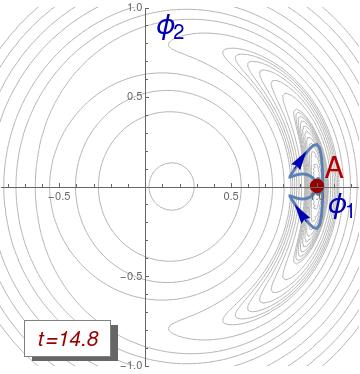}
			\includegraphics[height=2.cm]{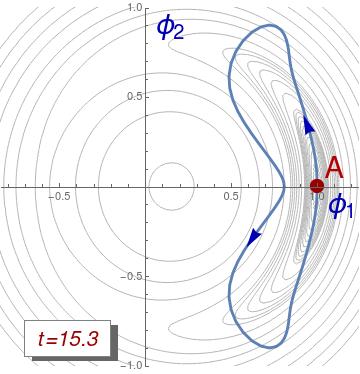}
			\includegraphics[height=2.cm]{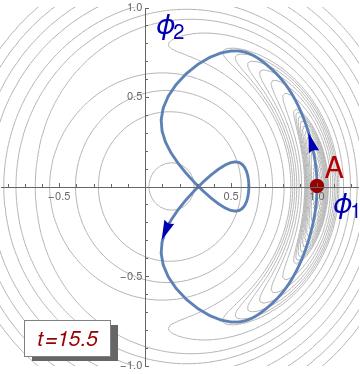}
			\includegraphics[height=2.cm]{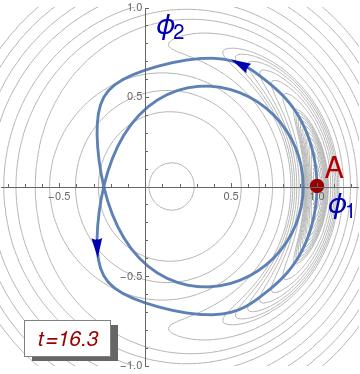}
			\includegraphics[height=2.cm]{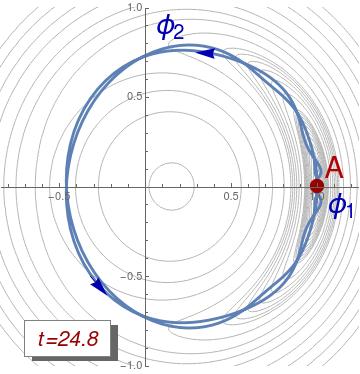} \end{tabular} }
	\caption{\small Evolution of the multi-kink orbit (\ref{conca2}) for a $K^{(w)}(x)-K^{(w)}(x)$-scattering process with impact velocity $v_0=0.68$ and model parameter $\sigma=0.95$ illustrated in Figure \ref{fig:flipreflection}. Arrows on the curves indicate the direction in which the kink orbit is traversed.}
	\label{fig:orbits22}
\end{figure}

	\item[\textbf{4.-}] \textit{Total annihilation:} The last type of possible final scenarios in this non-topological kink scattering context corresponds to the total annihilation of the two kinks. In this situation, the two kinks approach each other and collide; the strong fluctuations caused by the impact force the two loops of the evolving multi-kink orbit to jump the potential peak located at the point $P$. The radiation emission produced in this transition provokes a kinetic energy loss, which avoids that the kink trajectory could return to its original configuration. This means that only a radiation vestige (consisting of linear plane waves around the vacuum $A$) remains. This class of processes is symbolically represented as
	\[
	K^{(w)}(v_0) \cup K^{(w)}(-v_0) \rightarrow  \nu
	\]
	and has been illustrated in Figure \ref{fig:kinkkinkannihilation} for the case $\sigma=0.95$ with impact velocity $v_0=0.68$.
	
	\begin{figure}[h]
		\centerline{\includegraphics[height=3cm]{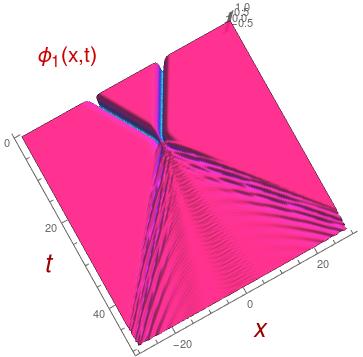} \hspace{1.5cm} \includegraphics[height=3cm]{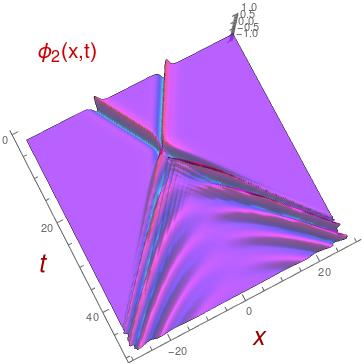}\hspace{1.5cm} \includegraphics[height=3cm]{colorbar}   }
		\caption{\small Evolution of the first and second scalar field components for a $K^{(w)}(x)$-$K^{(w)}(x)$ scattering process with impact velocity $v_0=0.7$ and model parameter $\sigma=2.0$.} \label{fig:kinkkinkannihilation}
	\end{figure}
\end{itemize}

\begin{figure}[h]
	\centerline{\begin{tabular}{c}
			\includegraphics[height=2.cm]{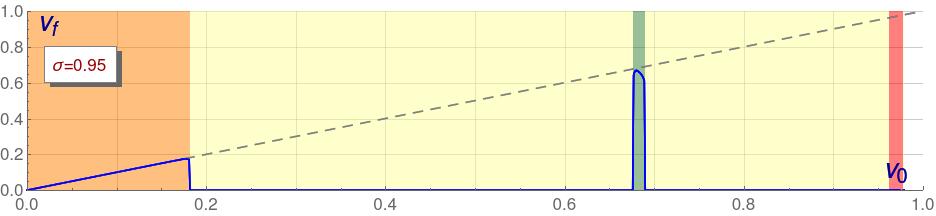}\\
			\includegraphics[height=2.cm]{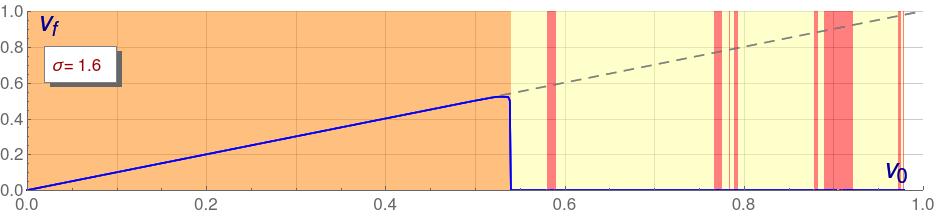}\\
			\includegraphics[height=2.cm]{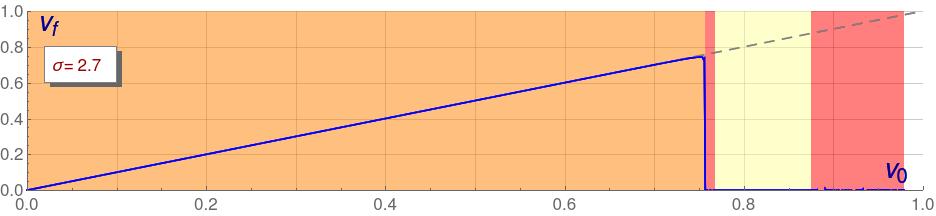}  \end{tabular} \\ \hspace{0.3cm}
		\begin{tabular}{c} \includegraphics[height=1.8cm]{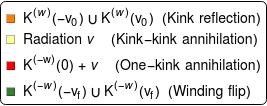}\end{tabular}}
	\caption{\small Graphical representation of the final velocity $v_f$ of the kinks as a function of the initial velocity $v_0$ for the ${K}^{(w)}(x)$-$K^{(w)}(x)$ scattering processes in the cases $\sigma=0.95$ (left, top), $\sigma=1.6$ (left, middle) and $\sigma=2.7$ (left, bottom). A color code distinguishing the final scenarios has been included (right).}
	\label{fig:diagrama1}
\end{figure}

In sum, four distinct scattering channels have been identified for the $K^{(w)}(x)-K^{(w)}(x)$ kink collisions: (1) kink reflection, (2) partial annihilation where globally one non-topological kink is destroyed, (3) kink reflection with winding number reversal and (4) total annihilation of the two kinks. All these scattering scenarios have been illustrated for particular values of the coupling constant $\sigma$ and the initial velocity $v_0$. The realization of these scattering events depends strongly on the impact velocity $v_0$ and the model parameter $\sigma$. In Figure \ref{fig:diagrama1} the final velocity $v_f$ of the scattered non-topological kinks is graphically represented as a function of the collision speed $v_0$ for several values of the coupling constant $\sigma$. For the sake of comparison a dashed straight line has been added in these figures representing the final velocity for elastic collisions. The four possible scattering channels have been distinguished by using a color stripe pattern. The color code (distinguishing the four possible final scenarios) has been included in Figure \ref{fig:diagrama1}. Obviously, when total annihilation (characterized by a yellow hue) occurs the final velocity is assumed to be zero. It is also noteworthy that for the one-kink (or partial) annihilation regime the surviving non-topological kink is pinned at the spatial axis origin, so the final velocity $v_f$ is also considered to vanish in this case. This situation is distinguished with a red hue from the previous case.

\noindent Some general characteristics can be recognized from Figure \ref{fig:diagrama1}, which are described as follows:

\vspace{0.1cm}

\noindent (1) Starting from small initial velocities $v_0$, kink reflection always emerges first. This regime is characterized by an orange hue in Figure \ref{fig:diagrama1}. We recall that in this case the two non-topological kinks collide and reflect each other in an almost elastic way. The initial velocity window associated with this regime gets broader as the coupling constant $\sigma$ increases. For example, the threshold values of this regime for the model parameters $\sigma=0.95$, $\sigma=1.6$ and $\sigma=2.7$ are  given by the impact velocities $v_0=0.182$, $v_0=0.540$ and $v_0=0.757$ respectively.

\vspace{0.1cm}

\noindent (2) The next predominant scattering regime corresponds to total annihilation, which can be distinguished by a yellow hue in Figure \ref{fig:diagrama1}. Here, the two kinks collide and are destroyed giving rise to a radiation vestige. As previously mentioned the final velocity in this regime is represented by a zero value. It can be checked that this regime takes up most of the diagram for the model parameter $\sigma=0.95$, whereas for the case $\sigma=2.7$ its presence is strongly reduced. This behavior is coherent with the fact that the potential barrier (which is encircled by the non-topological kink orbit) increases with the value of the coupling constant $\sigma$.   

\vspace{0.1cm}

\noindent (3) The one-kink (or partial) annihilation regime (displayed by a red hue in Figure \ref{fig:diagrama1}) follows a complex pattern. For these scattering events only one non-topological kink remains pinned at the spatial origin (with zero final velocity). For the model parameter $\sigma=0.95$, a small collision velocity window associated with this type of events arises for very high velocities. For $\sigma=1.6$, however, a sequence of bands with variable width are interspersed with other total annihilation windows. Finally, for the particular value $\sigma=2.7$ only two initial velocity windows for this regime appear nesting a total annihilation window. 

\vspace{0.1cm}

\noindent (4) Finally, for $\sigma=0.95$ a small window of collision velocities associated with the winding flip kink reflection has been found, see Figure \ref{fig:diagrama1}. Recall that in these cases when the kinks collide they reverse its winding charge.

\vspace{0.1cm}

\begin{figure}[h]
	\centerline{\begin{tabular}{cc} \includegraphics[height=6cm]{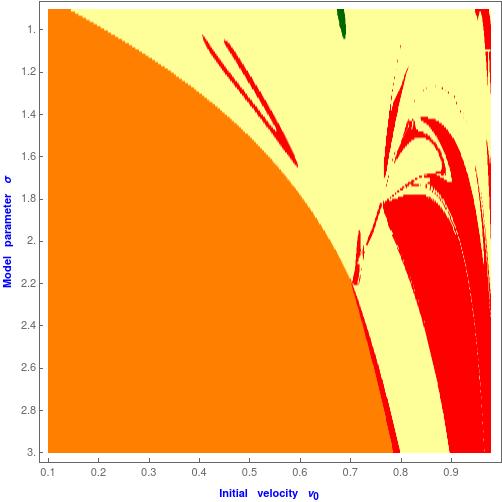}
	\end{tabular} \hspace{0.5cm} \begin{tabular}{c} \includegraphics[height=1.8cm]{leyenda11}
	\end{tabular}}
	\caption{\small Distribution of the different regimes for the $K^{(w)}(x)-K^{(w)}(x)$ kink scattering processes in the $(v_0,\sigma)$-plane for the range $\sigma\in [0.95,3]$ (left). A color code distinguishing the different regimes has been included (right). }
    \label{fig:GlobalDiagrama11}
\end{figure}

The results above reveal that the presence of the previously described regimes depends on the collision velocity $v_0$ and the coupling constant $\sigma$ in a non-trivial way. In order to understand this dependence, the distribution of the regimes in the $(v_0,\sigma)$-plane has been depicted in Figure \ref{fig:GlobalDiagrama11} for the range $\sigma\in [0.9,3.0]$. Kink scattering simulations for successive values of the coupling constant $\sigma$ with a step $\Delta \sigma = 0.01$ and an initial velocity step $\Delta v_0=0.001$ determines every pixel color in the graphics included in Figure \ref{fig:GlobalDiagrama11}. It can be observed that the elastic kink reflection regime grows as the coupling constant $\sigma$ increases. As previously mentioned the total annihilation regime predominates for small values of $\sigma$ in the interval $[0.9,3.0]$. Furthermore, a complex pattern emerges for the one-kink (partial) annihilation regime, which creates an astonishing graphic composition. Note the sequence of little red spots, which makes clear the existence of windows with very small width. For large enough values of $\sigma$ the total annihilation regime is confined between two one-kink annihilation bands.

\subsection{$K^{(w)}(x)$-$K^{(-w)}(x)$ kink scattering processes}

In this section the scattering between a non-topological kink $K^{(w)}(x)$ carrying winding charge $w$ and its own antikink $K^{(-w)}(x)$ with opposite winding charge is discussed. In other words, we shall study the collision between a basic extended particle and its antiparticle. The initial configuration (\ref{conca2}) employed in the numerical simulations for this case was described in Section 3.1 and illustrated in Figure \ref{fig:concatenation} (right). Both field components are even functions with respect to the space variable and will be so for any time instant because this parity symmetry is preserved by the Klein-Gordon equations (\ref{kleinequations}). Taking into account these conditions, the possible final scenarios in this context can be classified as follows:

\begin{itemize}
	\item[\textbf{1.-}] \textit{Bion formation:} In this type of events the non-topological kink and its antikink approach each other, collide and then pass through each other, moving away for a while until the attraction forces make them approach again. This pattern is repeated over and over forming a long-living kink-antikink bound state (known as bion). During every cycle a small amount of radiation is emitted  gradually decreasing its period. A different interpretation of these events is given as follows: the kink and its antikink approach each other, they collide and bounce back reversing its winding charges; after moving away for a while, they are obliged to approach again due to the attraction forces, colliding and bouncing back later while recovering its original winding charges. This behavior is indefinitely repeated. This class of scattering processes can be represented as
	\[
	K^{(w)}(v_0) \cup K^{(-w)}(-v_0) \rightarrow K^{(w)} \oplus K^{(-w)} + \nu
	\]
	where $K^{(w)} \oplus K^{(-w)}$ stands for the bion state. A particular bion formation event has been depicted in Figure \ref{fig:bionformation} for the particular values $\sigma=2.0$ and $v_0=0.02$.
	
	\begin{figure}[h]
		\centerline{\includegraphics[height=3cm]{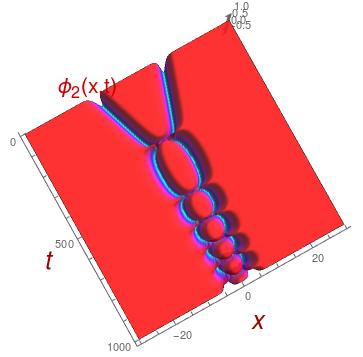} \hspace{1.5cm} \includegraphics[height=3cm]{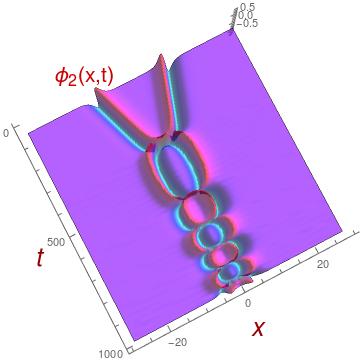}\hspace{1.5cm} \includegraphics[height=3cm]{colorbar}   }
		\caption{\small Evolution of the first and second scalar field components for a $K^{(w)}(x)$-$K^{(-w)}(x)$ scattering process with impact velocity $v_0=0.02$ and model parameter $\sigma=2.0$.} \label{fig:bionformation}
	\end{figure}

	\item[\textbf{2.-}] \textit{Kink-antikink passage:} In this occasion, the non-topological kink and its antikink approach each other, collide, pass through each other and move away with a final velocity approximately equal to the original collision velocity $v_0$. This behavior gives rise to the scattering process characterized by
	\[
	K^{(w)}(v_0) \cup K^{(-w)}(-v_0) \rightarrow K^{(-w)}(-v_f) \cup K^{(w)}(v_f) + \nu
	\]
	where radiation emission occurs if the initial velocity is large enough. An alternative interpretation is described as follows: kink and antikink approach each other, then they collide, reverse its winding charge and finally move away. In Figure \ref{fig:antireflection} this class of events has been graphically represented for the coupling constant $\sigma = 1.2$ and initial velocity $v_0=0.1$.
	
		\begin{figure}[h]
		\centerline{\includegraphics[height=3cm]{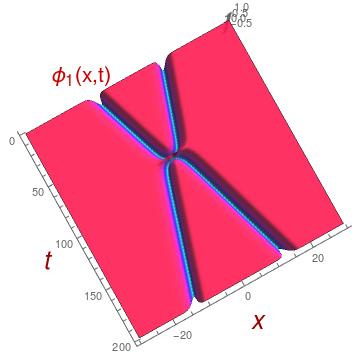} \hspace{1.5cm} \includegraphics[height=3cm]{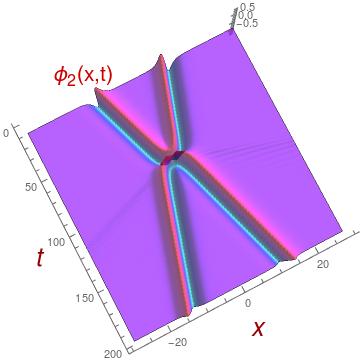}\hspace{1.5cm} \includegraphics[height=3cm]{colorbar}   }
		\caption{\small Evolution of the first and second scalar field components for a $K^{(w)}(x)$-$K^{(-w)}(x)$ scattering process with impact velocity $v_0=0.1$ and model parameter $\sigma=1.2$.} \label{fig:antireflection}
		\end{figure}

	\item[\textbf{3.-}] \textit{Kink-antikink annihilation:} Finally, a third final scenario can take place: kink and antikink approach each other and collide. After the impact kink and antikink emerge traveling away although the traversal fluctuations suffered in the collision, finally, make them decay to the vacuum solution. The global scattering process can be represented as
	\[
	K^{(w)}(v_0) \cup K^{(-w)}(-v_0) \rightarrow  \nu \hspace{0.5cm} .
	\]
	In Figure \ref{fig:highannihilation} a particular event of this type has been plotted for coupling constant $\sigma=1.2$ and impact velocity $v_0=0.9$.
	
		\begin{figure}[h]
		\centerline{\includegraphics[height=3cm]{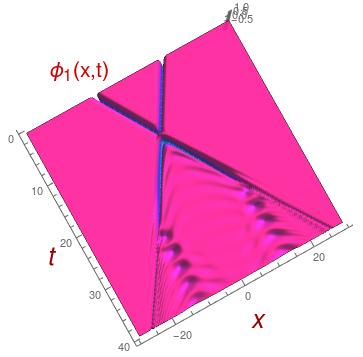} \hspace{1.5cm} \includegraphics[height=3cm]{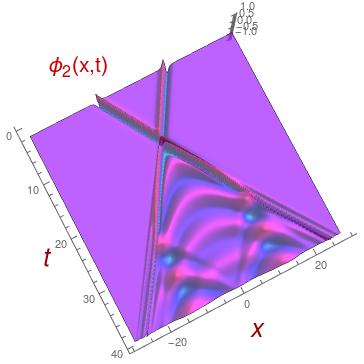} \hspace{1.5cm} \includegraphics[height=3cm]{colorbar}   }
		\caption{\small Evolution of the first and second scalar field components for a $K^{(w)}(x)$-$K^{(-w)}(x)$ scattering process with impact velocity $v_0=0.9$ and model parameter $\sigma=1.2$.} \label{fig:highannihilation}
		\end{figure}
	
\end{itemize}

\begin{figure}[h]
	\centerline{\begin{tabular}{c}
			\includegraphics[height=2.cm]{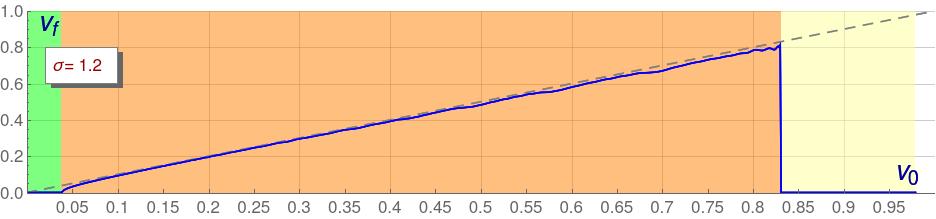}\\
			\includegraphics[height=2.cm]{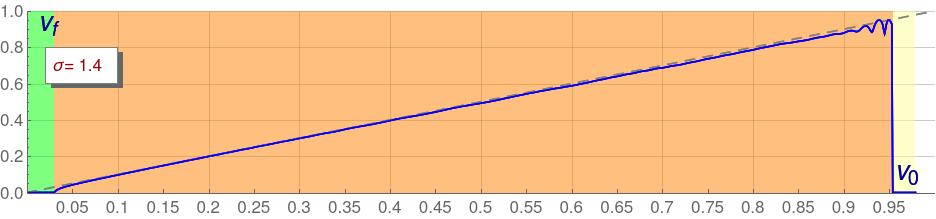}\\
			\includegraphics[height=2.cm]{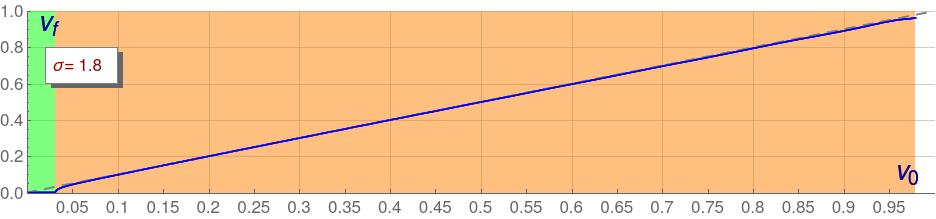}  \end{tabular} \\ \hspace{0.1cm}
		\begin{tabular}{c} \includegraphics[height=1.4cm]{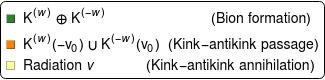}\end{tabular}}
	\caption{\small Graphical representation of the final velocity $v_f$ of the kinks as a function of the initial velocity $v_0$ for the ${K}^{(w)}(x)$-$K^{(-w)}(x)$ scattering processes in the cases $\sigma=1.20$ (left, top), $\sigma=1.40$ (left, middle) and $\sigma=1.80$ (left, bottom). A color code distinguishing the final scenarios has been included (right).}
	\label{fig:diagrama2}
\end{figure}

In sum, for the $K^{(w)}(x)$-$K^{(-w)}(x)$ kink scattering processes there exist three different scattering channels: (1) the bion formation, (2) the kink-antikink passage and (3) kink-antikink annihilation. These three types of events have been illustrated for particular values of the coupling constant and the initial velocity, see Figures \ref{fig:bionformation}, \ref{fig:antireflection} and \ref{fig:highannihilation} respectively. In Figure \ref{fig:diagrama2} the dependence of the final velocity $v_f$ of the scattered kinks on the initial velocity $v_0$  is depicted for the model parameters $\sigma=1.2$, $\sigma=1.4$ and $\sigma=1.8$. 

\begin{figure}[h]
	\centerline{ \begin{tabular}{c}\includegraphics[height=6cm]{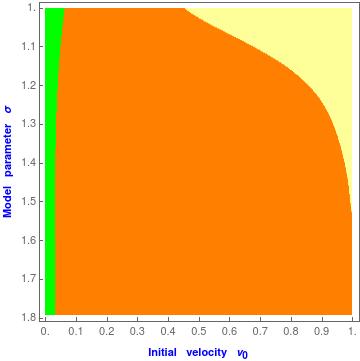}\end{tabular} \hspace{0.3cm} \begin{tabular}{c} \includegraphics[height=1.4cm]{leyenda12}\end{tabular} }
	\caption{\small Distribution of the different regimes for the $K^{(w)}(x)-K^{(-w)}(x)$ kink scattering processes in the $(v_0,\sigma)$-plane for the range $\sigma\in [1,2]$ (left). A color code distinguishing the different regimes has been included (right).}
	\label{fig:GlobalDiagrama12}
\end{figure}

It can be observed that the dominant behavior corresponds to the kink-antikink passage regime (represented by an orange hue in Figure \ref{fig:diagrama2}) where kink and antikink pass through each other in an almost elastic way. Exceptions to this rule are given for very low collision velocities where kink and antikink form a bound state corresponding to the bion formation regime (characterized by a green hue) and for very high impact velocities where kink and antikink decay to the vacuum solution giving rise to the annihilation regime (displayed by a yellow hue). The pattern found in the $K^{(w)}(x)$-$K^{(-w)}(x)$ kink scattering processes is much simpler than that found for the $K^{(w)}(x)$-$K^{(w)}(x)$ events, as can be checked in Figure \ref{fig:GlobalDiagrama12}, where the double dependence on the initial velocity $v_0$ and the coupling constant $\sigma$ of the different scattering channels is displayed. In addition, no resonant windows have been detected in this model.

\section{Conclussions}

In this paper the scattering between non-topological kinks arising in a two-component scalar field theory model has been analyzed. This model involves the presence of only one vacuum (absolute minima of the potential term $U(\phi)$), which implies that kink solutions must have a non-topological nature in this model. Despite of this fact, there exist stable non-topological kink solutions for the model parameter range $\sigma > 0.71$, whereas below such a threshold value the elastic forces lead to the kink collapse. Kink solutions can be interpreted as extended particles in a physical system. In this model these particles carry a winding charge $w=\pm 1$ because of the non-topological nature of the solutions. The scattering between non-topological kinks with the same and opposite winding charge has been considered in this work. In the first case, four different scattering channels have been identified, some of them corresponding to striking scattering events, such as the one-kink annihilation where the collision between two kinks leads to the annihilation of one half of each kink and the recombination of the other two halves, giving rise to a new non-topological kink with the opposite winding charge. The other three scattering channels in this framework correspond to the kink reflection, the winding flip kink reflection and the total annihilation. All of them have been described in Section 3.1. The presence of these scattering regimes depends non-trivially on the initial velocity $v_0$ and the coupling constant $\sigma$. This dependence leads to Figure \ref{fig:GlobalDiagrama11} where the distribution of the distinct regimes in the $(v_0,\sigma)$-plane is displayed. The second type of scattering events involves only three scattering channels: bion formation, kink-antikink passage and kink-antikink annihilation. The dependence in this case of the variables $v_0$ and $\sigma$ is much simpler than the previous case. Each regime occupies a well-distinguished region in the $(v_0,\sigma)$-plane, see Figure \ref{fig:GlobalDiagrama12}. 

As a final comment, it would be interesting to investigate kink dynamics within other two scalar field theory models in order to acquire a more global perspective of the problem. The study of kink dynamics in massive nonlinear $S^2$-sigma models \cite{Alonso2008, Alonso2009, Alonso2010, Alonso2018c} are natural candidates for carrying out this analysis and it also constitutes a challenging problem.

\section*{Acknowledgments}

The author acknowledges the Spanish Ministerio de Econom\'{\i}a y Competitividad for financial support under grant MTM2014-57129-C2-1-P. He is also grateful to the Junta de Castilla y Le\'on for financial help under grants VA057U16 and BU229P18. This research has made use of the high performance computing resources of the Castilla y Le\'on Supercomputing Center (SCAYLE, www.scayle.es), financed by the European Regional Development Fund (ERDF)

\end{document}